\documentclass[prl,aps,floats,superscriptaddress,floatfix,twocolumn,longbibliography]{revtex4-1}
\usepackage{amssymb,amsmath}
\usepackage{natbib}
\usepackage{amsmath,amssymb}
\usepackage{graphicx}
\usepackage{psfrag}
\usepackage{gensymb}

\usepackage{xcolor}
\usepackage[%
  colorlinks=true,
  urlcolor=blue,
  linkcolor=blue,
  citecolor=blue
]{hyperref}
\usepackage{etoolbox}

\usepackage{color}
\usepackage[normalem]{ulem}

\usepackage{float}
\restylefloat{table}
\usepackage{lipsum}

\usepackage{gensymb}

\newcommand{\br}{\mathbf{r}}

\newcommand{\bu}{\mathbf{u}}

\newcommand{\beq}{\begin{equation}}
\newcommand{\eeq}{\end{equation}}
\newcommand{\beqn}{\begin{eqnarray}}
\newcommand{\eeqn}{\end{eqnarray}}

\newcommand{\cO}{{\cal O}}

\def\beq{\begin{equation}}
\def\eeq{\end{equation}}
\def\bea{\begin{eqnarray}}
\def\eea{\end{eqnarray}}
\begin{document}
 \title{Rolled up or crumpled: phases of asymmetric tethered membranes}
\author{Tirthankar Banerjee}\email{tirthankar.banerjee@u-psud.fr}
\affiliation{LPTMS, UMR 8626, CNRS, Univ. Paris-Sud, Universit\'e Paris-Saclay, 91405 Orsay CEDEX, France}
\affiliation{Condensed Matter Physics Division, Saha Institute of
Nuclear Physics, 1/AF Bidhannagar, Calcutta 700064, West Bengal, India}
\author{Niladri Sarkar}\email{niladri2002in@gmail.com}
\affiliation{Max-Planck Institut f\"ur Physik Komplexer Systeme, N\"othnitzer 
Str. 38,
01187 Dresden, Germany}\affiliation{Laboratoire Physico
Chimie Curie, UMR 168, Institut Curie, PSL Research University,
CNRS, 
%75005 Paris, France.}\affiliation{
Sorbonne Universiti\'e,
%CNRS, Laboratoire Physico Chimie UMR 168 Curie,
 75005 Paris, France.}
\author{John Toner}\email{jjt@uoregon.edu}
\affiliation{Department of Physics and Institute of Theoretical Science, University of Oregon, Eugene, Oregon 97403, USA}
\author{Abhik Basu}\email{abhik.basu@saha.ac.in,abhik.123@gmail.com}
\affiliation{Condensed Matter Physics Division, Saha Institute of
Nuclear Physics, 1/AF Bidhannagar, Calcutta 700064, West Bengal, India} 
\affiliation{Max-Planck Institut f\"ur Physik Komplexer Systeme, N\"othnitzer 
Str. 38,
01187 Dresden, Germany}

\date{\today}
\begin{abstract}
  We show that inversion-asymmetric tethered membranes exhibit a new 
“double-spiral” phase with long range orientational order
not present in symmetric membranes. We calculate the universal algebraic spiral 
shapes of these
membranes in this phase. 
Asymmetry can trigger the crumpling of these membranes as well. 
 In-vitro experiments on  
lipid, 
red blood cell membrane extracts, 
and on graphene coated on one side,
could test these predictions.
\end{abstract}

\maketitle

%\section{Introduction}

%\label{intro}

The statistical mechanics of membranes has  
  long generated considerable 
theoretical and experimental interest~\cite{nelson2004}. 
In contrast to linear polymers~\cite{cates1984,cates1985}, fluctuating surfaces 
 can exhibit a wide variety of different phases, 
depending on rigidity, surface tension, and 
  topology.
Polymerized, or ``tethered" membranes
~\cite{nelson2004,kantor1987}
are 
two-dimensional (2D) analogs of linear polymer chains. But, unlike polymers, 
which are always coiled up, 
tethered membranes  
are known~\cite{nelson2004,paczuski1998} to 
display a statistically flat phase with long range orientational order (LRO) in 
the surface normals.
The very existence of a 
2D flat phase is surprising, since the well-known Hohenberg-Mermin-Wagner 
theorem (HMWT) forbids spontaneous symmetry breaking for 2D systems with a continuous symmetry~\cite{mermin1966,hohenberg1967}. 
Membranes get around this theorem via the coupling between 
in-plane elastic degrees of freedom and out-of-plane undulations,
 which
introduces an effective long-ranged interaction between those undulation modes. 

Most 
studies of tethered membranes
 have considered only
 {\it inversion-symmetric} membranes, i.e.,  
membranes that are identical 
on both sides. 
 Many real membranes, e.g., graphene coated on one side by some 
substance (e.g., polymer or a layer of lipid), {\em 
in-vivo} red blood cell membranes and {\em in-vitro} spectrin-deposited model 
lipid bilayers~\cite{lopez2012} are {\em structurally}  {\em inversion 
asymmetric}.
The effects of such asymmetry 
 are  still  largely unexplored.
 %\osut{  theoretically. }

\begin{figure}[htb]  
 \includegraphics[width=4cm]{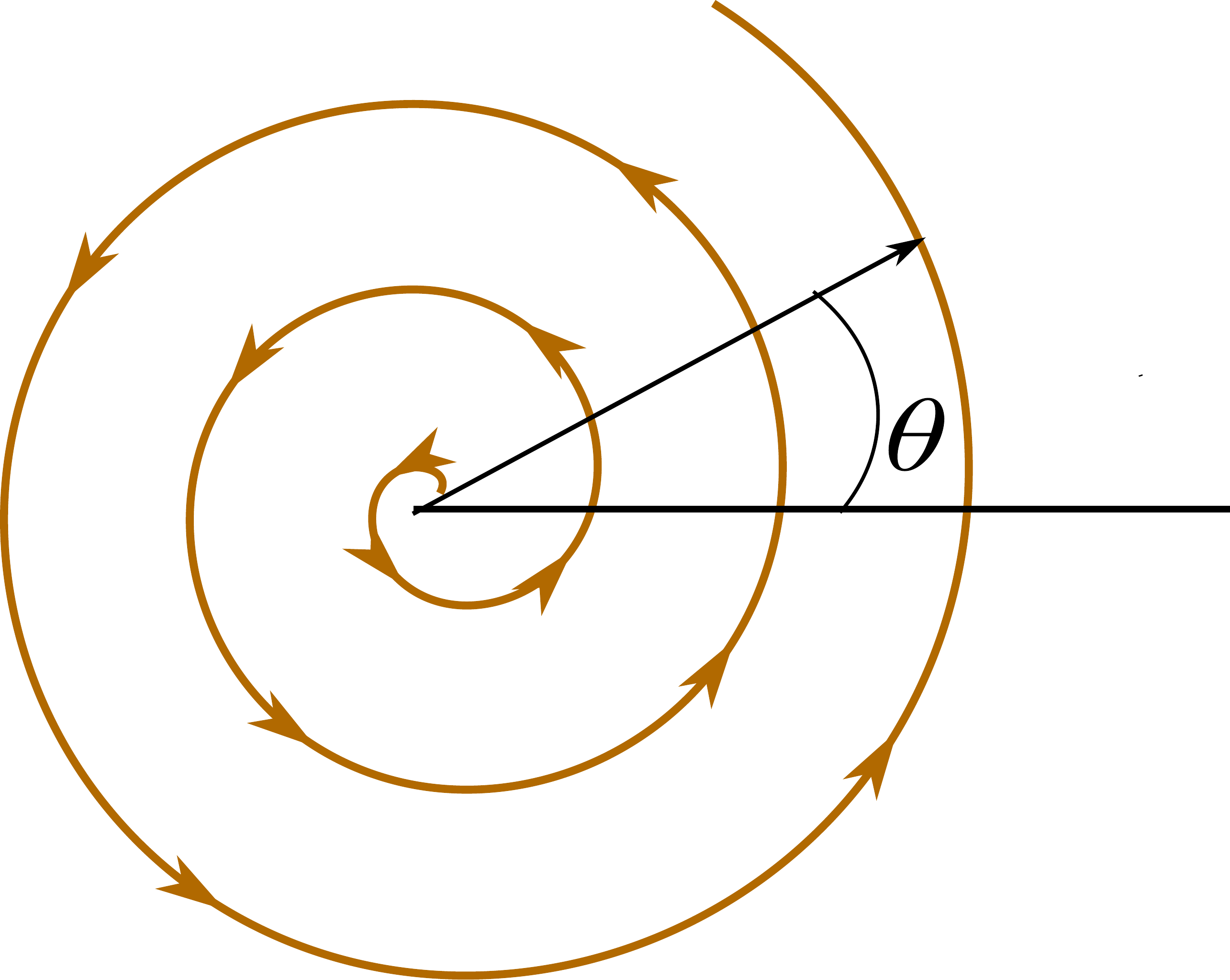}\hfill
 \includegraphics[width=4cm]{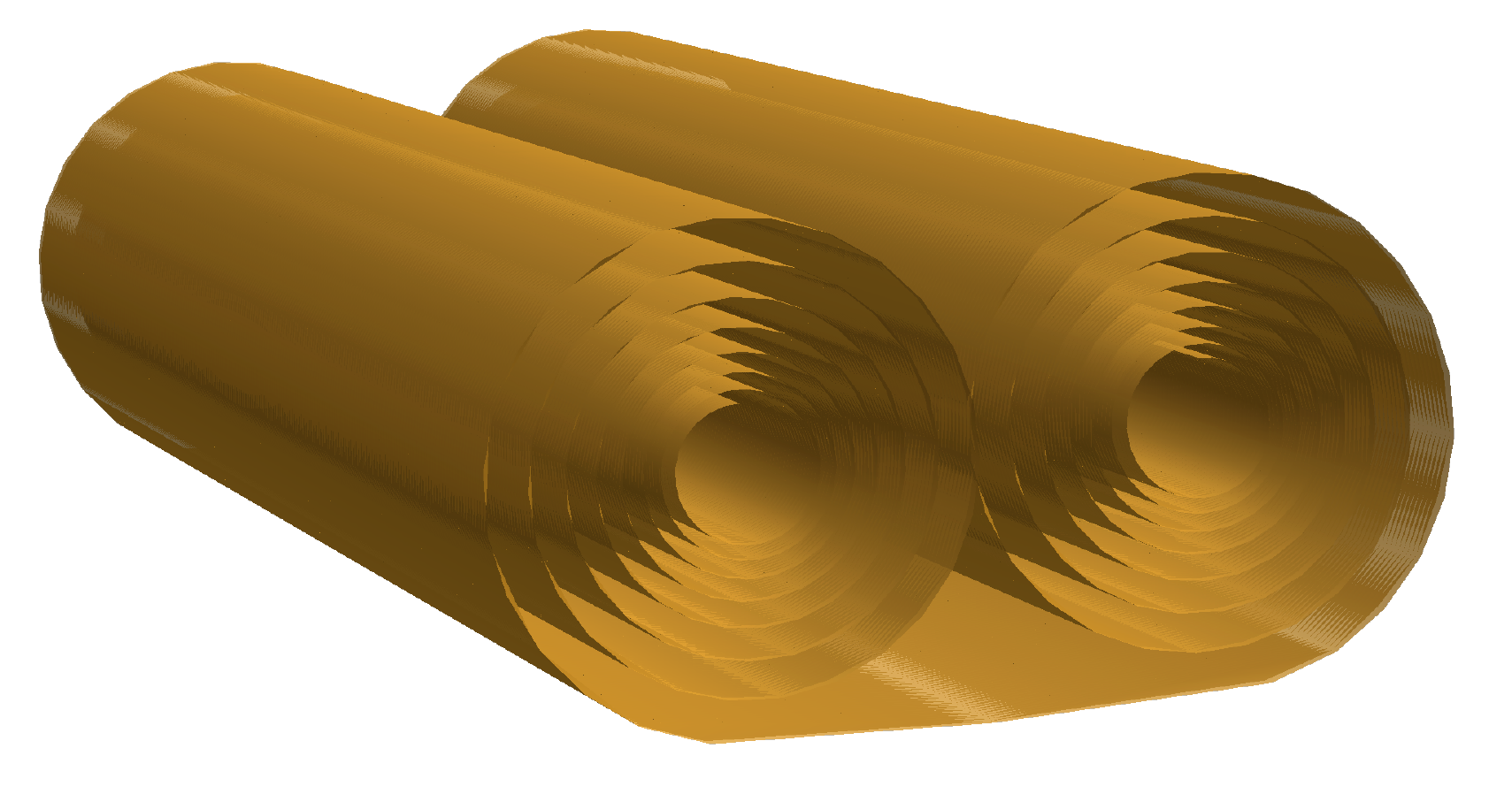}
 \caption{(Color online)(left) Schematic diagram of the cross section of 
  one of the double spirals, (right) Schematic diagram of a 
double spiral structure of 
 our model membrane.}
\label{dspiral}
 \end{figure}

In this Letter, we   investigate
the effects of asymmetry, and
 develop a generic and experimentally testable theory of 
equilibrium 
asymmetric tethered membranes (ATMs). 
We find that such membranes exhibit a new, 
"spiral state" not found in 
symmetric membranes.  As illustrated in 
Fig.~\ref{dspiral}, 
the mean spatial configuration of this state can be obtained by joining two 
coplanar spirals of opposite handedness at their base, and extruding that curve 
in the direction perpendicular to the plane of the spirals.   Note that this state, 
 like the flat phase of symmetric membranes, exhibits  long-ranged orientational order (LRO), although, obviously,  it has a very different structure.

 In the spiral state at temperature $T=0$, all ATMs  
assume the same shape.   This is also true at $T\ne0$, although the shape 
differs from that at $T=0$; in a sense, the $T\ne0$ shape is {\it infinitely} 
thermally expanded relative to the  $T=0$ configuration.
The $T=0$ configuration is, for  
large $L_m$,
simply a   double spiral of Archimedes,    each with a hole in the center; i.e., in polar coordinates,   each spiral is given by~%  JT: I cut this footnote to save space.}\sout{\footnote{Unsurprisingly, inversion-symmetric tethered membranes are always uncrumpled at $T=0$.}}:
\bea
r(\theta)=r_0 + a{\theta \over 2\pi} \ ; \label{rspiral}
\eea
see Fig.~\ref{dspiral} (right).
 In equation (\ref{rspiral}), $a$ is the 
thickness of the membrane,   $\theta=0$ corresponds to the inner edge of the membrane, and $r_0$ is the radius of the hole at the center of each spiral. Choosing this form 
for $r(\theta)$ simply means that the membrane is curled up as tightly as it can, 
given excluded volume effects.
 
 Thermal fluctuations considerably change this picture,   
  % \sout{open up  the spiral  by give rise 
%to a longer ranged "Helfrich repulsion"\cite{Helfrich} that }
opening up the spiral into the form:
 \bea
r(\theta)=R_0\theta^\beta, 
\label{rtheta} 
\eea 
where the universal exponent $\beta$ is related to the equally universal 
exponent $\eta$ characterizing the anomalous bend elasticity~\cite{nelson2004} of 
{\it symmetric} membranes through the relation 
$\beta={4 \over 2+\eta}{ 
\approx4-{2\over3}\sqrt{15}\approx1.418}$. 
The numerical estimate is 
based on the theoretical estimate
%\bea
$\eta=4/(1+\sqrt{15})\approx 0.821$
%\label{eta}\eea
 obtained by Radzihovsky and LeDoussal~\cite{Leo}. 
  In addition, the scale length $R_0$ exhibits universal scaling with 
temperature and  membrane parameters, which can also be related exactly to the 
exponent $\eta$; in particular, we find
\bea
R_0\propto T^{2(2-\eta){/(2+\eta)}}{\approx T^{0.836} }\,.
\label{R_01}
\eea

The total radius 
$R_T$ of the spiral regions also exhibits universal scaling,  in 
this case  with
the spatial extent $L_m$ of the membrane:
\bea
R_T={ R_0^{1-\alpha}} L_m ^\alpha \,\,\,\,,\,\,\,\, \alpha\equiv{4\over6+\eta}\approx 0.586\,\,.
\label{ralpha} 
\eea

%\bea\Gamma\propto T^{(2-\eta)} \,.\label{Gamma}\eea

Increasing asymmetry eventually induces
a novel structural instability 
which
actually 
crumples  the membrane~\cite{peliti1985,nelson2004}.
In further contrast with symmetric membranes, 
we find two distinct regimes of parameter space  within the crumpled phase of  
ATMs. In one  of these, (hereafter called ``strongly crumpled" , or ``SC"), 
 the membrane will be crumpled no matter how small it is, while in the second 
(hereafter called "weakly crumpled" or ``WC"), it is only crumpled if 
its lateral spatial extent $L_m$ exceeds a critical size $L_c$, which 
depends on material parameters of the membrane. Smaller 
membranes (i.e., $L_m<L_c$) exhibit a spiral structure similar to that found in 
the spiral phase, but different in its scaling properties.  Those scaling properties can be obtained from those specified by equations (\ref{rtheta}), (\ref{R_01}) and (\ref{ralpha}) by replacing $\eta$ everywhere it appears by $0$.
This  crumpling behavior is 
summarized in Fig.~\ref{k-chi},  in which $\chi$ is a phenomenological parameter (defined in equation (\ref{free energy}) below) characterizing the asymmetry of the membrane, with $\chi=0$ for symmetrical membranes.

   \begin{figure}[htb]  
 \includegraphics[width=5.5cm,height=4cm]{phase2.pdf}\hfill 
\includegraphics[width=6.5cm]{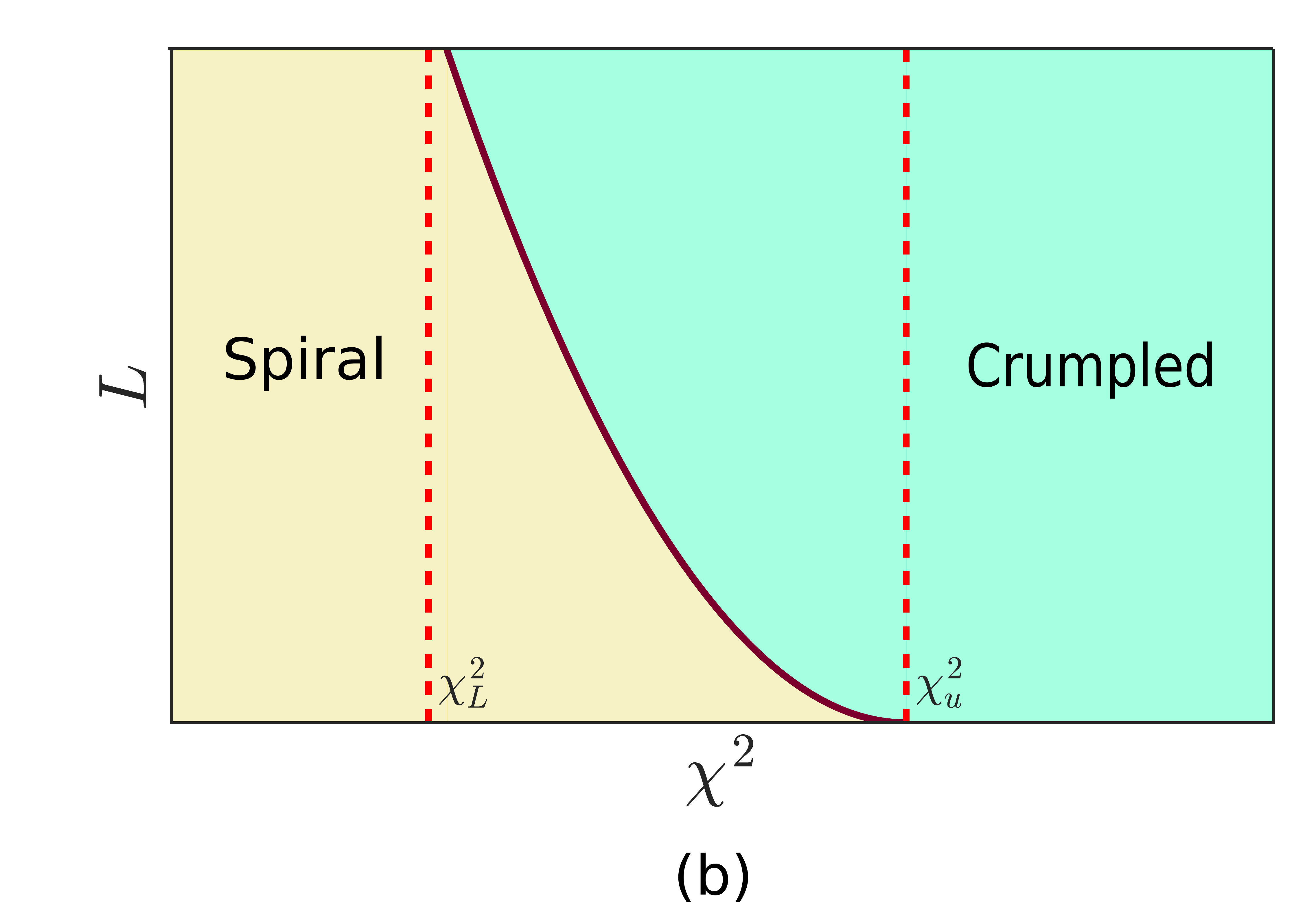}
 \caption{(Color online)
(a) Schematic 
``phase diagram" in the $\chi^2-\kappa'$ plane. (b) Schematic 
 ``phase diagram" in the $\chi^2-L$ plane for fixed $\kappa$. The 
continuous curve (black) is the line $L=\xi (\chi^2)$, demarcating the  spiral
and crumpled phases. }
\label{k-chi}
 \end{figure}

 We will  now outline the derivation of these results; more detail is 
given in the  associated long paper (ALP)~\cite{banerjee}. 

We begin by 
formulating the elastic model for a single turn of the spiral structure, on 
length scales short compared to both the local radius of curvature  
$R$ and the
typical distance $L_H$ between successive  points of contact between that turn  and
the turns immediately
inside and outside of it.  A membrane  patch of linear size $L\ll L_H$ 
behaves 
like an isolated, free membrane  with no contact with 
anything else. The results 
of this analysis will then be used as 
inputs to treat the membrane on progressively larger scales: first, to compute 
$L_H$, and thereby calculate the interaction between successive turns of 
the membrane, and then on length scales comparable to   $R$, 
to calculate the large scale  spiral structure of the membrane. 

 For $L\ll L_H$ and 
$L\ll  R$, %\sout{  where $ R$ is} the radius of curvature,} 
we can
describe the membrane 
 by a single-valued field $h({\bf r})$  in the Monge gauge and 
lateral displacement by a 2D vector field ${\bf u}({\bf 
r})$~\cite{nelson2004,chaikin2000}.
General symmetry considerations then 
dictate  the following  form
 for the free energy functional $F$ for 
 elastically isotropic (i.e., either amorphous or hexagonal crystalline), 
tensionless asymmetric tethered membranes:
\begin{eqnarray}
F&=&\frac{1}{2}\int d^2r \left[\kappa'(\nabla^2 h)^2+ 
\lambda u_{ii}^2 +
2\mu u_{ij}u_{ij} + 2\chi u_{ii}\nabla^2 h \right]\nonumber \\ &+&\int 
d^2r\,\,C\nabla^2 h
\label{free energy}
\end{eqnarray}
to 
leading order in gradients; here $r=|{\bf 
r}|$, ${\bf r}=(x,y)$ with (${\bf 
r}, h$) denoting the coordinate of a point on the membrane in the 
three-dimensional embedding space.  The strain tensor $u_{ij}=\frac{1}{2}(\nabla_i u_j + 
\nabla_j u_i +\nabla_i h 
\nabla_j h)$, 
 ignoring irrelevant terms, and $\nabla^2 h$ is approximately the mean 
curvature for nearly flat membranes.
 This model~(\ref{free energy}) 
 differs from the model for symmetric 
membranes~\cite{chaikin2000,aronovitz1988,nelson2004} by the addition of  two 
generic 
inversion-symmetry breaking terms:  a linear
``spontaneous curvature" term $C\nabla^2 h$, that 
makes the membrane want to curl up
with a radius of 
curvature $R_{ s}\propto 1/C$
and a  term  $\chi u_{ii}\nabla^2 h $, that  favors local 
 bending of the 
membrane in response to local compression of the elastic network. See Ref.~\cite{leibler1986} for a term analogous to our $\chi$ term introduced for  {\it fluid} membranes.

 Working to quadratic order in the fields, and integrating over $\bu$,  gives
an effective 
 free energy functional that depends only on $h({\bf r})$: 
\beq
{\mathcal F}_{eff} = \int \frac{d^2 r}{(2\pi)^2} \left[\frac{\kappa_{ 0}}{2} 
(\nabla^2 h 
)^2+C\nabla^2 h\right] \ ,
\label{Fquadh}
\eeq 
with an 
effective bending modulus
 $\kappa$: 
 \begin{equation}\label{kappa}
  \kappa_0=\kappa'-\frac{\chi^2}{2\mu+\lambda}{ \equiv\kappa'(1-\chi^2/\chi_{_U}^2)} \,,
 \end{equation}

\noindent { where we have defined}
\begin{equation}
\chi_U^2=\kappa'(2\mu+\lambda) \,.
\label{chiu}
 \end{equation}
Evidently, $\kappa_0<\kappa'$. 
Thermodynamic stability 
of the 
  membrane clearly requires $\kappa_0>0$, otherwise 
  instability ensues. This implies { that $\chi_{_U}^2$ is } an instability threshold for $\chi^2$, with larger $\chi^2$'s being unstable. 
      This is the 
asymmetry-induced 
crumpling discussed earlier in this Letter, which we see can occur even at 
$T=0$. Since (\ref{kappa})   is 
$q$-independent, 
any ensuing crumpling takes place at {\em all scales}, meaning  an 
arbitrarily small membrane 
 will be crumpled, provided $\kappa_0<0$.
We will see later that anharmonic effects actually cause the membrane to 
crumple for a larger range of $\chi$'s; specifically, when 
$\chi^2>\chi_L^2$, although for $\chi_L^2<\chi^2<\chi_U^2$, 
crumpling only occurs if the membrane is sufficiently large.

%Reorganize:  1st,  consider structure when $\kappa>0$ and $T=0$. Here, we show 
%this structure is a spiral. Then turn to finite $T$.

We now turn to the uncrumpled case $\kappa_{0}>0$, 
and show that the ground state structure is the double spiral of Archimedes 
illustrated in Fig.~\ref{dspiral}. 
Since ${\mathcal F}$
 is bilinear in $u_{i}$, we can 
 follow \cite{nelson2004} and integrate {\em exactly} over $u_{i}$ 
 in calculating the partition function associated with (\ref{free 
energy}) to arrive at an effective free energy $F_h$
 that depends only on $h$ (now including anharmonic terms in $h$).
  The result, given in detail in the ALP, is a model with the same 
long ranged interaction between Gaussian curvatures  as in   symmetric
tethered membranes~\cite{nelson2004}, and a   new, weaker, 
but still long-ranged, interaction between Gaussian and mean curvature $\nabla^2 h$   that is unique to asymmetric membranes,
%  JT cut an equation here}
%\begin{widetext}  \begin{eqnarray}  \label{free energy_h}  {\mathcal F}_h = \frac{1}{2}\int d^2 r \left[ \kappa_0(\nabla^2 h)^2+2C\nabla^2 h\right]  +\frac{1}{2}\int d^2 r\, d^2 r'\left[ {\cblue A} U_G(\br-\br')G(\br)G(\br')  + B U_{GM}(\br-\br')(\nabla^2 h(\br))G(\br')\right] , \end{eqnarray} \end{widetext}}
  where the local Gaussian  curvature is given approximately by 
$G(\br)\approx(\partial_x^2h) (\partial_y^2h)-(\partial_x\partial_yh)^2$   for nearly flat membranes. 
 This long-ranged interaction between   Gaussian curvatures $G(\br)$ at different points suppresses Gaussian curvature, causing the membrane to curl in only one direction in the spiral state.
 %  AB: Niladri and Tirthankar had some confusions on what ``This'' in the previous line referred to. So I added a small text.}

%one direction. 
In the ALP, we show
that 
the long-ranged interactions between the mean and the Gaussian curvature do not alter this conclusion.
%\sout{ This is why the "spiral" state of the membrane curls in only one direction. }

We will now more precisely determine the shape of the membrane at $T=0$. 
Minimizing (\ref{Fquadh}) over the mean inverse radius of curvature $\nabla^2 h$  
implies that the membrane energetically prefers to curl up with 
$\nabla^2 h={1\over R_1}=-{C\over\kappa_0}\equiv -{1\over R_s}$, where we have 
defined the spontaneous radius of curvature $R_s={\kappa_0\over C}$.  However, a membrane of lateral extent $L>{\pi R_s^2\over a}$, 
where $a$ is the thickness of the membrane, cannot fit into a 
cylinder of radius $R_s$, because its total volume 
$L^2a$ will be greater than the volume of a cylinder of radius $R_s$ and length $L$. 
Therefore, the best the membrane can do is to curl up as tightly as it can without overlapping itself. 
The shape that accomplishes this while bending in only one direction is 
the double spiral of Archimedes described by (\ref{rspiral}).
The  reason {\it two} spirals form is that 
by so doing the membrane can reduce the 
average value of $R_1$, since each spiral only has to wind out 
${1\over\sqrt{2}}$ as far.

\begin{figure}[htb] 
 \includegraphics[width=7cm]{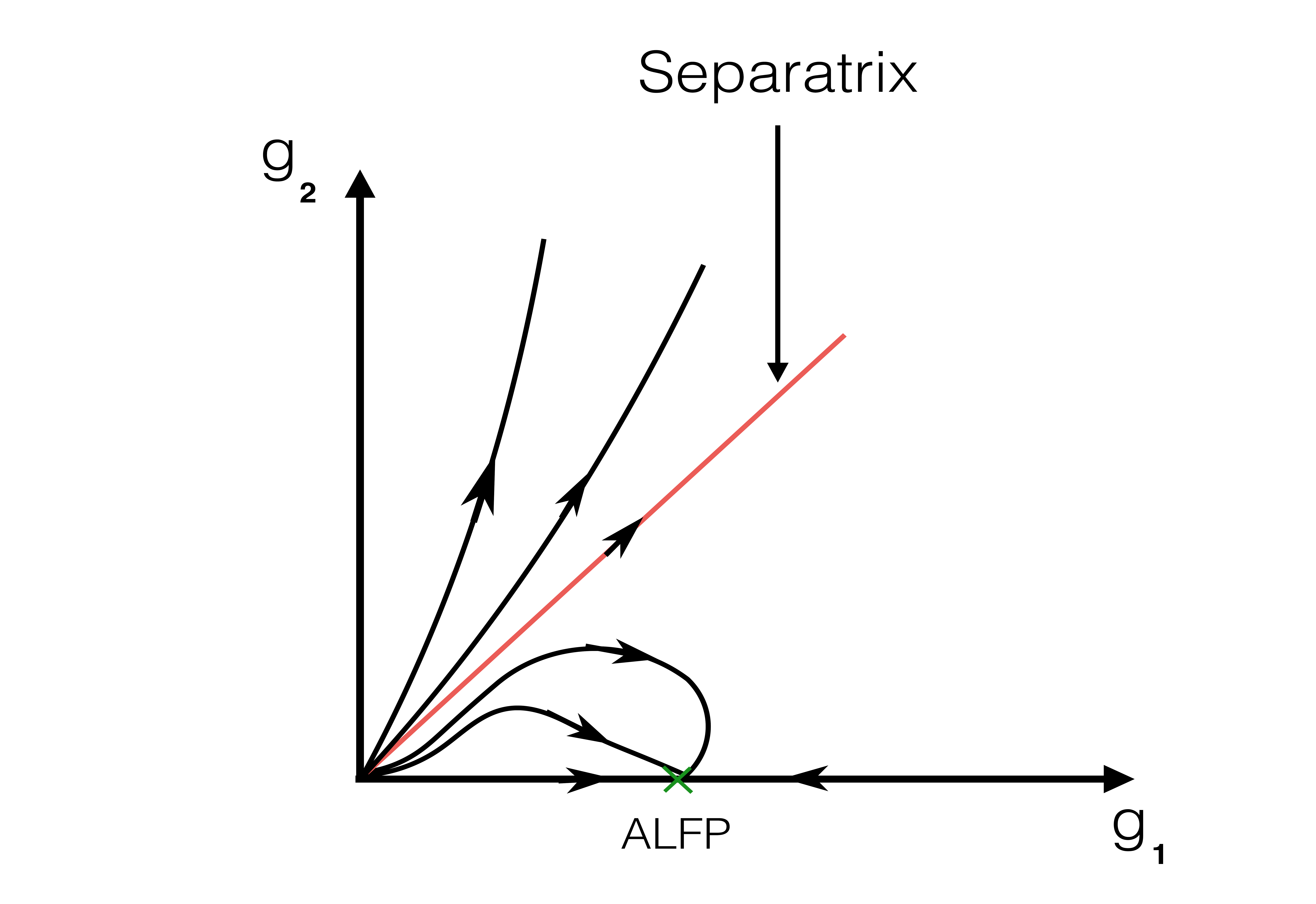} 
\caption{(Color online) Schematic flow lines in the $g_1-g_2$ plane. 
(Green) cross marks the stable FP ($2\epsilon/5,0$). The red straight line is 
the 
separatrix  $g_2=3g_1/5$. } 
\label{fp}
 \end{figure}

As in symmetric membranes~\cite{aronovitz1988, chaikin2000, nelson2004},  
 at non-zero temperatures, the combination of thermal fluctuations and 
anharmonic effects 
substantially modify the behavior of the membrane. 
To study this, we perform a perturbative renormalization group (RG) analysis of 
the model (\ref{free energy}),  which we remind the reader is only valid on length scales $L\ll L_H$. 
As usual,   the RG is done by
 tracing over the short wavelength Fourier modes
  of  $h({\bf r})$, followed by a 
rescaling of lengths and 
$h$. This  leads to the following
differential 
recursion relations:
\begin{eqnarray}
   \frac{d\kappa}{dl}=\kappa\left[-\eta + g_1
   - \frac{5}{2}g_2\right]\,,
   \label{renor_kappa}\\
 \frac{dg_1}{dl}=g_1\left[\epsilon -\frac{5g_1}{2}+5g_2\right] \,,\label{flow1}\\
 \frac{dg_2}{dl}=g_2\left[\epsilon-4g_1+\frac{15}{2}g_2\right] \,,\label{flow2}
%\frac{dC}{dl}=(2-\epsilon+\eta)C + g_2 \times O(1)\,,\label{flowc}
\end{eqnarray} 
where  $\kappa(l=0)=\kappa_0$, and we have defined two effective coupling 
constants, 
\beq
g_1\equiv\frac{AS_Dk_BT\Lambda^{-\epsilon}}{(2\pi)^D\kappa^2}\,\,\,,\,\,\,\,
g_2\equiv\frac{B^2S_Dk_BT\Lambda^{-\epsilon}}{(2\pi)^D 
\kappa^3}\,\,,
\label{gdef}
\eeq
with
$A\equiv\frac{4\mu(\mu+\lambda)}{2\mu+\lambda}>0$ and 
$B\equiv\frac{2\chi\mu}{2\mu+\lambda}$. Here,
$\exp (l)$ is   the length  rescaling 
factor, 
$\epsilon\equiv4-D$, where D is the "internal"  dimension 
of 
the membrane (e.g., D=2 in the physical case), and  
$S_D$ is the surface area  of a 
D-dimensional sphere of unit radius.
 The flows in the $g_1$-$g_2$ plane are illustrated in Fig.~\ref{fp}.

% {\bf I didn't follow the last line. It's true that C doesn't get any diagrammatic correction, but C(l) changes due to the naive rescaling of space. 

%{\cred [JT: Yes, but when you actually use the RG to compute the actual curvature of the membrane in the original, unrescaled coordinates, all of those effects that come from rescaling cancel out (as they should, since they are just changes of variable that don't reflect any physics).]}

%Or did you mean, the unscaled C is length-scale independent? 

%{\cred JT: Yes; exactly! (Genau!)]}

%In that case, in what sense this is the physical C? 

%{\cred [JT In the sense that if you balance the physical $\kappa$ against the physical $ C$ defined in this way, you get the actual radius of curvature of the actual membrane, in the original unrescaled coordinates.]}

%Is this the one that will be measured in exp?} 

%{\cred [JT: You bet!]}

As can been seen from that figure, the only stable fixed point is at $g_2=0$, $g_1=2\epsilon/5\,$,  which we will denote by call ``ALFP"  (Aronovitz-Lubensky fixed point) hereafter~\cite{aronovitz1988}. Since
 only $g_2$  (which depends on $B$, and hence $\chi$) and 
$C$ involve any parameter  that breaks the up-down symmetry of the lattice, 
 the vanishing of $g_2$ at the fixed point implies that the large scale 
properties of any system whose starting parameters lie in the basin of 
attraction of this fixed
point (i.e., the region below the separatrix in Fig.~\ref{fp}) will, with 
increasing length scale,  become identical to those of a symmetric membrane,  
until we reach length scales at which $C$ becomes 
important.  Membranes in this region of parameter space 
are in the spiral state discussed below, whereas membranes whose 
bare $g_{1,2}$ lie above the separatrix $g_2=3g_1/5$ are crumpled, as  we 
will 
discuss later.

We now turn to the effects of thermal fluctuations on the spiral 
phase itself. This requires studying the system at larger scales $L\gg 
L_H$.  We still expect a 
double spiral (see Fig.~\ref{dspiral}) when $T>0$.  
 Thermal fluctuations  open up the spiral 
 by giving rise to a longer ranged "Helfrich repulsion" 
 $\mathcal{U}_{H}(d)$~\cite{helfrich1978,toner1990} that 
 is caused by
excluded volume interactions  between parts of neighboring turns of the 
membrane that have made large excursions from their mean position (here $d$  is 
the local mean separation between successive turns~\footnote{Here by ``local mean'', we mean the separation averaged over a ``local'' region small enough to look like a stack of parallel, flat membranes; this means the averaging volume must have a linear extent much less that the local radius of curvature, but much greater than the distance $L_H$ between successive membrane contacts.}. This interaction  
 has the same form and 
scaling as for a lamellar phase of {\it symmetric} membranes\cite{toner1990}, since $g_2\to0$ 
upon renomalization below the separatrix, as discussed above. 
As shown in detail in the ALP, the result of balancing this interaction against 
the spontaneous curvature energy is the form of the spiral given by equations  
(\ref{rtheta}),  (\ref{R_01}),  and (\ref{ralpha}).
Note that 
this spiral 
phase displays orientational LRO.

 Having discussed the spiral phase, 
we turn now  to the other
region of parameter space, namely that  which flows away 
from the ALFP, and towards negative $\kappa$. 
While 
we cannot follow these flows all the way to $\kappa
=0$ (since both $g_{1,2}$ diverge there, so that our perturbation theory breaks down), we suspect that this signals crumpling of 
large membranes. This region of parameter space  therefore corresponds to the 
crumpled phase. 
For $\epsilon=4-D\ll1$, which is the region in which our perturbative RG is 
accurate, this is the region in figure (\ref{k-chi}) lying
above the separatrix  $g_2=3g_1/5$. 
For the physical case $\epsilon=2$, as discussed in the ALP, it seems 
reasonable to assume that there continues to be a separatrix which, for small 
$g_{1,2}$, is a straight line 
%\beq
$g_2=\rho g_1$
%\eeq
of universal slope $\rho=\cO(1)$, although since $\epsilon=2$ we cannot
calculate the universal constant $\rho$. 

The range of
$\chi$ in our original model (\ref{free energy}) that we are now discussing is 
$\chi_L^2<\chi^2<\chi_U^2$, where the upper bound follows because we are 
considering positive $\kappa _0$ in Eq.~(\ref{kappa}), while the lower bound follows 
from assuming that we are above the separatrix, which implies, for small bare 
$g^0_{1,2}$, that $g^0_2/g^0_1>\rho $. Using our earlier expressions (\ref{gdef}) for 
$g_{1,2}$, we see that this implies
\beq
\chi^2>{\rho\kappa'(2\mu_0+\lambda_0)\over\rho+{\mu_0\over\mu_0+\lambda_0}}
={\chi_U^2\over 1+{\mu_0\over\rho(\mu_0+\lambda_0)}}\equiv\chi_L^2 \,,
\label{chiL3}
\eeq
where in the equality we have used our result (\ref{chiu}) for $\chi_u^2$.
Note that, reassuringly, we always have $\chi_L^2<\chi_U^2$, since $\rho$, $\mu_0$, and $\mu_0+\lambda_0$ are all positive, the latter two positivities being required for stability.

For $\chi$'s in the range $\chi_L^2<\chi^2<\chi_U^2$, 
the membrane can remain  uncrumpled if it is sufficiently small. 
This is because, in this  range of $\chi$'s, the 
bare value 
$\kappa(\ell=0)=\kappa_0=\kappa'-\frac{\chi^2}{2\mu_0+\lambda_0}$,  is positive, and can stabilize 
orientational order and thereby prevent crumpling. Hence, that order 
will only be lost on length scales  $L>\xi$, where $\xi$ is the smallest length 
scale  big enough to allow enough renormalization group ``time" $\ell$ for 
$\kappa(\ell)$ to be driven to zero. 
This implies that 
the membrane will crumple unless new physics beyond the purely elastic 
model (\ref{free energy}) intervenes on some length scale  smaller than $\xi$.
%{\cred JT cut an equation here}
%\beq \xi=\Lambda^{-1}e^{\ell_0} \,,\label{xi1}\eeq

%}
%{\cred JT cut an equation here}
%\beq
%\xi=\Lambda^{-1}e^{\ell_0} \,,
%\label{xi1}
%\eeq

%{\cred A detailed calculation, given in the ALP, shows that }

Now we need to consider what "new physics"  
beyond the elastic model (\ref{free energy}) can intervene  before this length 
scale is reached to prevent crumpling. 
 One possibility is self-avoidance, which can cut off any tendency to crumpling in 
the spiral sections of the membrane. But as inspection of figure (\ref{dspiral}) 
makes clear, %
this cut off cannot work for the straight section connecting the two oppositely returning spirals.  This section 
has no neighbors, because it lies outside both spirals. It is 
therefore the section of the membrane that will crumple first, thereby inducing 
crumpling of the rest of the membrane. 

This straight, ``connecting" section of the membrane is stabilized by  surface tension, which arises because that section of the membrane could lower its energy by `rolling up" into one or the other of the spiral sections it connects. 
It is not rolled up, of course, because the other spiral 
pulls it equally hard in the opposite direction. These two pulls create a non-zero surface tension $\sigma$, whose magnitude should be comparable to the Helfrich interaction in outermost turn of spiral, since it is the balance between that interaction, which works to open the spiral, and the spontaneous curvature term, which tries to tighten it, that sets the scale of that spontaneous curvature energy, and, hence, the surface tension. In the ALP we  use  
this reasoning 
to calculate this surface tension $\sigma$, and the associated length scale  $L_\sigma=\sqrt{\kappa\over\sigma}$ obtained by equating $\sigma$ to the bending energy ${\kappa_0\over L_\sigma^2} $.
%\beq L_\sigma=\sqrt{\kappa\over\sigma} \sim\left({k_BT\kappa_0L_m\over C_0^2}\right)^{1/3}\ .\label{Lsig}\eeq
Equating  $L_\sigma$ to $\xi$ and solving for $L_m$ gives the maximum size 
$L_c$ of the membrane that can be stable:
\beq 
L_c\sim{\kappa_0^{7/2}C_0^2\over (k_BT)^{5/2}B_0^3}\propto (\chi_U^2 - \chi^2)^{7/2}  \,,\label{L_c}
\eeq
where the final proportionality follows from our expression  (\ref{kappa}) for $\kappa_0$.
See Figs.~\ref{k-chi} for schematic phase diagrams in 
the $\chi-\kappa'$ and $\chi^2-L$ planes. 
%{\cred JT notes: we can't really say anything for $\chi\approx\chi_U$, since $g_{1,2}\to\infty$ there (Because $\kappa\to0$ there). Fortunately, for small $\kbt$, we can get very close to $\chi_U$ before pert theory breaks down (because $g_{1,2}\propto\kbt$). The above argument can probably get the right ``phase boundary" in figure 3 in that regime.}

%\beq L_c\sim{\kappa_0^{7/2}C_0^2\over (k_BT)^{5/2}B_0^3}\propto (\chi_U^2 - \chi^2)^{7/2}  \,.\label{L_c}\eeq
 This scaling law breaks down both near $\chi_U$, where $L_c$ gets to be  $R_s$, so the membrane is not long enough to wind up at all, and as $\chi\to\chi_L$,  for reasons discussed in the ALP.

 %{\bf Lastly, ... we need the T-dependence of $\xi$.}

%\section{Model asymmetric tethered membrane}
%\label{modelmem}

%\begin{figure}[htb]
%%\includegraphics[width=8cm]{rbcpic.pdf}
%\caption{(color online) Schematic top view of a membrane (black quadrangle) 
%coupled to an
%elastic network (broken red triangular lattice) on the bottom side, $l_0$ is 
%the average distance between the two.  The membrane and the 
%elastic network are joined at the lattice points (see, e.g., 
%Ref.~\cite{alberts2014} for RBC membrane structures, not shown in this 
%diagram); see 
%text.  }  \label{modelrbc}
%\end{figure}

\noindent {\it Summary:} We have developed the theory of asymmetric 
tethered membranes. This theory predicts a spiral state, with
the shape of the membrane  
at $T=0$ a double Archimedes spiral at $T=0$,
and an algebraic spiral with a universal exponent at $T>0$. 
We also find that sufficiently  asymmetric membranes are crumpled; the mechanism for this is quite different from ``buckling" of elastic shells \cite{kovsmrlj2017}.  This leads to the  phase 
diagrams 
(\ref{k-chi}),  which
   can be tested in non living (ATP-depleted) RBC membrane 
extract~\cite{lopez2012}, 
 for model asymmetric membranes by binding spectrin to lipids 
~\cite{o2000}, or graphene coated with some substance (e.g., polymer or a 
layer of lipid) on one side, as well as by numerical 
simulations\cite{abraham89,kantor1987,peng2013}. We  
hope our work   will stimulate
experimental and numerical studies of asymmetric tethered membranes.

\paragraph{Acknowledgements:-}
T.B. and A.B.  thank the Alexander von
Humboldt Stiftung (Germany) for partial financial support under the Research 
Group Linkage Programme scheme (2016). 
T.B. and J. T. thank the Max-Planck Institut f\"ur Physik Komplexer Systeme, 
Dresden, Germany, for their hospitality and financial support while this
 work was underway.

\bibliography{tethered}

\begin{thebibliography}{21}%
\makeatletter
\providecommand \@ifxundefined [1]{%
 \@ifx{#1\undefined}
}%
\providecommand \@ifnum [1]{%
 \ifnum #1\expandafter \@firstoftwo
 \else \expandafter \@secondoftwo
 \fi
}%
\providecommand \@ifx [1]{%
 \ifx #1\expandafter \@firstoftwo
 \else \expandafter \@secondoftwo
 \fi
}%
%
\providecommand \enquote  [1]{``#1''}%
\providecommand \bibnamefont  [1]{#1}%
\providecommand \bibfnamefont [1]{#1}%
\providecommand \citenamefont [1]{#1}%
\providecommand \href@noop [0]{\@secondoftwo}%
\providecommand \href [0]{\begingroup \@sanitize@url \@href}%
\providecommand \@href[1]{\@@startlink{#1}\@@href}%
\providecommand \@@href[1]{\endgroup#1\@@endlink}%
\providecommand \@sanitize@url [0]{\catcode `\\12\catcode `\$12\catcode
  `\&12\catcode `\#12\catcode `\^12\catcode `\_12\catcode `\%12\relax}%
\providecommand \@@startlink[1]{}%
\providecommand \@@endlink[0]{}%
\providecommand \url  [0]{\begingroup\@sanitize@url \@url }%
\providecommand \@url [1]{\endgroup\@href {#1}{\urlprefix }}%
\providecommand \urlprefix  [0]{URL }%
%
%
\providecommand \selectlanguage [0]{\@gobble}%
\providecommand \bibinfo  [0]{\@secondoftwo}%
\providecommand \bibfield  [0]{\@secondoftwo}%
%
\providecommand \BibitemOpen [0]{}%
%
%
%
\providecommand \BibitemShut  [1]{\csname bibitem#1\endcsname}%
\let\auto@bib@innerbib\@empty
%</preamble>
\bibitem [{\citenamefont {Nelson}\ \emph {et~al.}(2004)\citenamefont {Nelson},
  \citenamefont {Piran},\ and\ \citenamefont {Weinberg}}]{nelson2004}%
  \BibitemOpen
  \bibfield  {author} {\bibinfo {author} {\bibfnamefont {D.~R.}\ \bibnamefont
  {Nelson}}, \bibinfo {author} {\bibfnamefont {T.}~\bibnamefont {Piran}}, \
  and\ \bibinfo {author} {\bibfnamefont {S.}~\bibnamefont {Weinberg}},\
  }\href@noop {} {\emph {\bibinfo {title} {Statistical mechanics of membranes
  and surfaces}}}\ (\bibinfo  {publisher} {World Scientific},\ \bibinfo {year}
  {2004})\BibitemShut {NoStop}%
\bibitem [{\citenamefont {Cates}(1984)}]{cates1984}%
  \BibitemOpen
  \bibfield  {author} {\bibinfo {author} {\bibfnamefont {M.~E.}\ \bibnamefont
  {Cates}},\ }\bibfield  {title} {\enquote {\bibinfo {title} {Statics and
  dynamics of polymeric fractals},}\ }\href@noop {} {\bibfield  {journal}
  {\bibinfo  {journal} {Phys. Rev. Lett.}\ }\textbf {\bibinfo {volume} {53}},\
  \bibinfo {pages} {926--929} (\bibinfo {year} {1984})}\BibitemShut {NoStop}%
\bibitem [{\citenamefont {Cates}(1985)}]{cates1985}%
  \BibitemOpen
  \bibfield  {author} {\bibinfo {author} {\bibfnamefont {M.~E.}\ \bibnamefont
  {Cates}},\ }\bibfield  {title} {\enquote {\bibinfo {title} {The fractal
  dimension and connectivity of random surfaces},}\ }\href@noop {} {\bibfield
  {journal} {\bibinfo  {journal} {Phys. Lett. B}\ }\textbf {\bibinfo {volume}
  {161}},\ \bibinfo {pages} {363--367} (\bibinfo {year} {1985})}\BibitemShut
  {NoStop}%
\bibitem [{\citenamefont {Kantor}\ \emph {et~al.}(1987)\citenamefont {Kantor},
  \citenamefont {Kardar},\ and\ \citenamefont {Nelson}}]{kantor1987}%
  \BibitemOpen
  \bibfield  {author} {\bibinfo {author} {\bibfnamefont {Y.}~\bibnamefont
  {Kantor}}, \bibinfo {author} {\bibfnamefont {M.}~\bibnamefont {Kardar}}, \
  and\ \bibinfo {author} {\bibfnamefont {D.~R.}\ \bibnamefont {Nelson}},\
  }\bibfield  {title} {\enquote {\bibinfo {title} {Tethered surfaces: Statics
  and dynamics},}\ }\href@noop {} {\bibfield  {journal} {\bibinfo  {journal}
  {Phys. Rev. A}\ }\textbf {\bibinfo {volume} {35}},\ \bibinfo {pages} {3056}
  (\bibinfo {year} {1987})}\BibitemShut {NoStop}%
\bibitem [{\citenamefont {Paczuski}\ \emph {et~al.}(1988)\citenamefont
  {Paczuski}, \citenamefont {Kardar},\ and\ \citenamefont
  {Nelson}}]{paczuski1998}%
  \BibitemOpen
  \bibfield  {author} {\bibinfo {author} {\bibfnamefont {M.}~\bibnamefont
  {Paczuski}}, \bibinfo {author} {\bibfnamefont {M.}~\bibnamefont {Kardar}}, \
  and\ \bibinfo {author} {\bibfnamefont {D.~R.}\ \bibnamefont {Nelson}},\
  }\bibfield  {title} {\enquote {\bibinfo {title} {Landau theory of the
  crumpling transition},}\ }\href@noop {} {\bibfield  {journal} {\bibinfo
  {journal} {Phys. Rev. Lett.}\ }\textbf {\bibinfo {volume} {60}},\ \bibinfo
  {pages} {2638--2640} (\bibinfo {year} {1988})}\BibitemShut {NoStop}%
\bibitem [{\citenamefont {Mermin}\ and\ \citenamefont
  {Wagner}(1966)}]{mermin1966}%
  \BibitemOpen
  \bibfield  {author} {\bibinfo {author} {\bibfnamefont {N.~D.}\ \bibnamefont
  {Mermin}}\ and\ \bibinfo {author} {\bibfnamefont {H.}~\bibnamefont
  {Wagner}},\ }\bibfield  {title} {\enquote {\bibinfo {title} {Absence of
  ferromagnetism or antiferromagnetism in one-or two-dimensional isotropic
  heisenberg models},}\ }\href@noop {} {\bibfield  {journal} {\bibinfo
  {journal} {Phys. Rev. Lett.}\ }\textbf {\bibinfo {volume} {17}},\ \bibinfo
  {pages} {1133} (\bibinfo {year} {1966})}\BibitemShut {NoStop}%
\bibitem [{\citenamefont {Hohenberg}(1967)}]{hohenberg1967}%
  \BibitemOpen
  \bibfield  {author} {\bibinfo {author} {\bibfnamefont {P.~C.}\ \bibnamefont
  {Hohenberg}},\ }\bibfield  {title} {\enquote {\bibinfo {title} {Existence of
  long-range order in one and two dimensions},}\ }\href@noop {} {\bibfield
  {journal} {\bibinfo  {journal} {Phys. Rev.}\ }\textbf {\bibinfo {volume}
  {158}},\ \bibinfo {pages} {383} (\bibinfo {year} {1967})}\BibitemShut
  {NoStop}%
\bibitem [{\citenamefont {L{\'o}pez-Montero}\ \emph {et~al.}(2012)\citenamefont
  {L{\'o}pez-Montero}, \citenamefont {Rodr{\'\i}guez-Garc{\'\i}a},\ and\
  \citenamefont {Monroy}}]{lopez2012}%
  \BibitemOpen
  \bibfield  {author} {\bibinfo {author} {\bibfnamefont {I.}~\bibnamefont
  {L{\'o}pez-Montero}}, \bibinfo {author} {\bibfnamefont {R.}~\bibnamefont
  {Rodr{\'\i}guez-Garc{\'\i}a}}, \ and\ \bibinfo {author} {\bibfnamefont
  {F.}~\bibnamefont {Monroy}},\ }\bibfield  {title} {\enquote {\bibinfo {title}
  {Artificial spectrin shells reconstituted on giant vesicles},}\ }\href@noop
  {} {\bibfield  {journal} {\bibinfo  {journal} {J. Phys. Chem. Lett.}\
  }\textbf {\bibinfo {volume} {3}},\ \bibinfo {pages} {1583--1588} (\bibinfo
  {year} {2012})}\BibitemShut {NoStop}%
\bibitem [{\citenamefont {Doussal}\ and\ \citenamefont
  {Radzihovsky}(1992)}]{Leo}%
  \BibitemOpen
  \bibfield  {author} {\bibinfo {author} {\bibfnamefont {P.~Le}\ \bibnamefont
  {Doussal}}\ and\ \bibinfo {author} {\bibfnamefont {L.}~\bibnamefont
  {Radzihovsky}},\ }\bibfield  {title} {\enquote {\bibinfo {title}
  {Self-consistent theory of polymerized membranes},}\ }\href@noop {}
  {\bibfield  {journal} {\bibinfo  {journal} {Phys. Rev. Lett.}\ }\textbf
  {\bibinfo {volume} {69}},\ \bibinfo {pages} {1209} (\bibinfo {year}
  {1992})}\BibitemShut {NoStop}%
\bibitem [{\citenamefont {Peliti}\ and\ \citenamefont
  {Leibler}(1985)}]{peliti1985}%
  \BibitemOpen
  \bibfield  {author} {\bibinfo {author} {\bibfnamefont {L.}~\bibnamefont
  {Peliti}}\ and\ \bibinfo {author} {\bibfnamefont {S.}~\bibnamefont
  {Leibler}},\ }\bibfield  {title} {\enquote {\bibinfo {title} {Effects of
  thermal fluctuations on systems with small surface tension},}\ }\href@noop {}
  {\bibfield  {journal} {\bibinfo  {journal} {Phys. Rev. Lett.}\ }\textbf
  {\bibinfo {volume} {54}},\ \bibinfo {pages} {1690} (\bibinfo {year}
  {1985})}\BibitemShut {NoStop}%
\bibitem [{\citenamefont {Banerjee}\ \emph {et~al.}(2018)\citenamefont
  {Banerjee}, \citenamefont {Sarkar}, \citenamefont {Toner},\ and\
  \citenamefont {Basu}}]{banerjee}%
  \BibitemOpen
  \bibfield  {author} {\bibinfo {author} {\bibfnamefont {T.}~\bibnamefont
  {Banerjee}}, \bibinfo {author} {\bibfnamefont {N.}~\bibnamefont {Sarkar}},
  \bibinfo {author} {\bibfnamefont {J.}~\bibnamefont {Toner}}, \ and\ \bibinfo
  {author} {\bibfnamefont {A.}~\bibnamefont {Basu}},\ }\bibfield  {title}
  {\enquote {\bibinfo {title} {Statistical mechanics of asymmetric tethered
  membranes: spiral and crumpled phases},}\ }\href@noop {} {\bibfield
  {journal} {\bibinfo  {journal} {companion long paper}\ } (\bibinfo {year}
  {2018})}\BibitemShut {NoStop}%
\bibitem [{\citenamefont {Chaikin}\ and\ \citenamefont
  {Lubensky}(2000)}]{chaikin2000}%
  \BibitemOpen
  \bibfield  {author} {\bibinfo {author} {\bibfnamefont {P.~M.}\ \bibnamefont
  {Chaikin}}\ and\ \bibinfo {author} {\bibfnamefont {T.~C.}\ \bibnamefont
  {Lubensky}},\ }\href@noop {} {\emph {\bibinfo {title} {Principles of
  condensed matter physics}}}\ (\bibinfo  {publisher} {Cambridge university
  press},\ \bibinfo {year} {2000})\BibitemShut {NoStop}%
\bibitem [{\citenamefont {Aronovitz}\ and\ \citenamefont
  {Lubensky}(1988)}]{aronovitz1988}%
  \BibitemOpen
  \bibfield  {author} {\bibinfo {author} {\bibfnamefont {J.~A.}\ \bibnamefont
  {Aronovitz}}\ and\ \bibinfo {author} {\bibfnamefont {T.~C.}\ \bibnamefont
  {Lubensky}},\ }\bibfield  {title} {\enquote {\bibinfo {title} {Fluctuations
  of solid membranes},}\ }\href@noop {} {\bibfield  {journal} {\bibinfo
  {journal} {Phys. Rev. Lett.}\ }\textbf {\bibinfo {volume} {60}},\ \bibinfo
  {pages} {2634} (\bibinfo {year} {1988})}\BibitemShut {NoStop}%
\bibitem [{\citenamefont {Leibler}(1986)}]{leibler1986}%
  \BibitemOpen
  \bibfield  {author} {\bibinfo {author} {\bibfnamefont {S.}~\bibnamefont
  {Leibler}},\ }\bibfield  {title} {\enquote {\bibinfo {title} {Curvature
  instability in membranes},}\ }\href@noop {} {\bibfield  {journal} {\bibinfo
  {journal} {J. Physique}\ }\textbf {\bibinfo {volume} {47}},\ \bibinfo {pages}
  {507--516} (\bibinfo {year} {1986})}\BibitemShut {NoStop}%
\bibitem [{\citenamefont {Helfrich}(1978)}]{helfrich1978}%
  \BibitemOpen
  \bibfield  {author} {\bibinfo {author} {\bibfnamefont {W.}~\bibnamefont
  {Helfrich}},\ }\bibfield  {title} {\enquote {\bibinfo {title} {Steric
  interaction of fluid membranes in multilayer systems},}\ }\href@noop {}
  {\bibfield  {journal} {\bibinfo  {journal} {Z. Naturforsch. A}\ }\textbf
  {\bibinfo {volume} {33}},\ \bibinfo {pages} {305--315} (\bibinfo {year}
  {1978})}\BibitemShut {NoStop}%
\bibitem [{\citenamefont {Toner}(1990)}]{toner1990}%
  \BibitemOpen
  \bibfield  {author} {\bibinfo {author} {\bibfnamefont {J.}~\bibnamefont
  {Toner}},\ }\bibfield  {title} {\enquote {\bibinfo {title} {New phase of
  matter in lamellar phases of tethered, crystalline membranes},}\ }\href@noop
  {} {\bibfield  {journal} {\bibinfo  {journal} {Phys. Rev. Lett.}\ }\textbf
  {\bibinfo {volume} {64}},\ \bibinfo {pages} {1741} (\bibinfo {year}
  {1990})}\BibitemShut {NoStop}%
\bibitem [{Note1()}]{Note1}%
  \BibitemOpen
  \bibinfo {note} {Here by ``local mean'', we mean the separation averaged over
  a ``local'' region small enough to look like a stack of parallel, flat
  membranes; this means the averaging volume must have a linear extent much
  less that the local radius of curvature, but much greater than the distance
  $L_H$ between successive membrane contacts.}\BibitemShut {Stop}%
\bibitem [{\citenamefont {Ko{\v{s}}mrlj}\ and\ \citenamefont
  {Nelson}(2017)}]{kovsmrlj2017}%
  \BibitemOpen
  \bibfield  {author} {\bibinfo {author} {\bibfnamefont {A.}~\bibnamefont
  {Ko{\v{s}}mrlj}}\ and\ \bibinfo {author} {\bibfnamefont {D.~R.}\ \bibnamefont
  {Nelson}},\ }\bibfield  {title} {\enquote {\bibinfo {title} {Statistical
  mechanics of thin spherical shells},}\ }\href@noop {} {\bibfield  {journal}
  {\bibinfo  {journal} {Phys. Rev. X}\ }\textbf {\bibinfo {volume} {7}},\
  \bibinfo {pages} {011002} (\bibinfo {year} {2017})}\BibitemShut {NoStop}%
\bibitem [{\citenamefont {O'Toole}\ \emph {et~al.}(2000)\citenamefont
  {O'Toole}, \citenamefont {Morrison},\ and\ \citenamefont {Cherry}}]{o2000}%
  \BibitemOpen
  \bibfield  {author} {\bibinfo {author} {\bibfnamefont {P.~J.}\ \bibnamefont
  {O'Toole}}, \bibinfo {author} {\bibfnamefont {I.~E.~G.}\ \bibnamefont
  {Morrison}}, \ and\ \bibinfo {author} {\bibfnamefont {R.~J.}\ \bibnamefont
  {Cherry}},\ }\bibfield  {title} {\enquote {\bibinfo {title} {Investigations
  of spectrin--lipid interactions using fluoresceinphosphatidylethanolamine as
  a membrane probe},}\ }\href@noop {} {\bibfield  {journal} {\bibinfo
  {journal} {Biochimica et Biophysica Acta (BBA)-Biomembranes}\ }\textbf
  {\bibinfo {volume} {1466}},\ \bibinfo {pages} {39--46} (\bibinfo {year}
  {2000})}\BibitemShut {NoStop}%
\bibitem [{\citenamefont {Abraham}\ \emph {et~al.}(1989)\citenamefont
  {Abraham}, \citenamefont {Rudge},\ and\ \citenamefont
  {Plischke}}]{abraham89}%
  \BibitemOpen
  \bibfield  {author} {\bibinfo {author} {\bibfnamefont {F.~F.}\ \bibnamefont
  {Abraham}}, \bibinfo {author} {\bibfnamefont {W.~E.}\ \bibnamefont {Rudge}},
  \ and\ \bibinfo {author} {\bibfnamefont {M.}~\bibnamefont {Plischke}},\
  }\bibfield  {title} {\enquote {\bibinfo {title} {Molecular dynamics of
  tethered membranes},}\ }\href@noop {} {\bibfield  {journal} {\bibinfo
  {journal} {Phys. Rev. Lett.}\ }\textbf {\bibinfo {volume} {62}},\ \bibinfo
  {pages} {1757} (\bibinfo {year} {1989})}\BibitemShut {NoStop}%
\bibitem [{\citenamefont {Peng}\ \emph {et~al.}(2013)\citenamefont {Peng},
  \citenamefont {Li}, \citenamefont {Pivkin}, \citenamefont {Dao},
  \citenamefont {Karniadakis},\ and\ \citenamefont {Suresh}}]{peng2013}%
  \BibitemOpen
  \bibfield  {author} {\bibinfo {author} {\bibfnamefont {Z.}~\bibnamefont
  {Peng}}, \bibinfo {author} {\bibfnamefont {X.}~\bibnamefont {Li}}, \bibinfo
  {author} {\bibfnamefont {I.~V.}\ \bibnamefont {Pivkin}}, \bibinfo {author}
  {\bibfnamefont {M.}~\bibnamefont {Dao}}, \bibinfo {author} {\bibfnamefont
  {G.~E.}\ \bibnamefont {Karniadakis}}, \ and\ \bibinfo {author} {\bibfnamefont
  {S.}~\bibnamefont {Suresh}},\ }\bibfield  {title} {\enquote {\bibinfo {title}
  {Lipid bilayer and cytoskeletal interactions in a red blood cell},}\
  }\href@noop {} {\bibfield  {journal} {\bibinfo  {journal} {Proc. Nat. Acad.
  Sc. (USA)}\ }\textbf {\bibinfo {volume} {110}},\ \bibinfo {pages}
  {13356--13361} (\bibinfo {year} {2013})}\BibitemShut {NoStop}%
\end{thebibliography}

\end{document}